\documentclass{proclund}

\newcommand{\Oeepp}{\mbox{$^{16}$O(e,e$'$pp)~}}
\newcommand{\Oeenp}{\mbox{$^{16}$O(e,e$'$np)~}}

\newcommand{\eep}{\mbox{(e,e$'$p)~}}
\newcommand{\eenp}{\mbox{(e,e$'$np)~}}
\newcommand{\eepp}{\mbox{(e,e$'$pp)~}}
\newcommand{\eex}{\mbox{(e,e$'$X)}}
\newcommand{\Oeep}{\mbox{$^{16}$O(e,e$'$p)~}}
\newcommand{\OeepN}{\mbox{$^{16}$O(e,e$'$p)$^{15}$N~}}

\newcommand{\deep}{\mbox{$^{2}$H(e,e$'$p)n~}}
\newcommand{\Aeep}{\mbox{A(e,e$'$p)~}}
\newcommand{\AeepB}{\mbox{A(e,e$'$p)A$-$1~}}

\begin{document}
\pagestyle{prochead}


\title{WHAT DO WE LEARN FROM \eep EXPERIMENTS}
\author{S. Gilad}
\email{sgilad@mitlns.mit.edu}
\affiliation
 {Massachusetts Institute of Technology, Cambridge, MA, USA\\~\\}

\begin{abstract}
The promise and limitations of studying nuclear structure and reactions
using \eep reactions are discussed.  \AeepB reactions on complex nuclei 
are used to test new relativistic mean-field calculations and their
ingredients such as ground-state wave function, current operators,
final-state interactions and two-body currents, and to assess
the limits of single-particle models.  For higher excitations
of the residual nuclei, \Aeep reactions reveal only qualitative
features of the reaction mechanism, and their increased usefulness  
awaits input from more complex reactions such as A\eex.  The \eep
experiments on few-body nuclei provide essential input to quickly
developing theories for few-body systems, where individual theoretical
ingredients can be studied very selectively.  A large data set
is needed for this purpose, and existing or proposed experimental
programs are discussed.  If hadronic theories can meet the challenge 
of consistently describing this body of data at low and
moderate energies, the transition from hadronic to quark-gluon
degrees of freedom could be probed in experiments at higher energies.
\end{abstract}

\maketitle

\setcounter{page}{1}

\section{Introduction}

\begin{figure}[htb]
    \includegraphics[height=0.25\textheight]{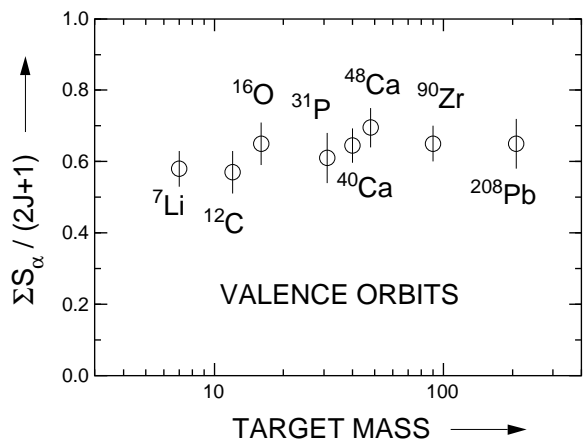}
    \caption{\label{fig:sfa}
      Spectroscopic factors for valence states of different nuclei.
      These spectroscopic factors were extracted using the independent
      particle shell model (IPSM)~\cite{lapikas1}.}
\end{figure}

Exclusive \AeepB reactions have been used in the last two decades to
study nuclear structure.  Cross-sections have been measured for specific
states of the residual nuclei as a function of the missing momentum,
$p_\mathrm{m}$.  The electron-nucleon cross-section was then divided out
of the measured cross-sections to extract the ``distorted'' momentum
distribution for a specific hole state. The data, mainly for low
transfered momenta ($Q^2\le 0.4$ (GeV/c)$^2$), have been analyzed
in terms of non-relativistic 
model-dependent DWIA calculations. The final-state interactions (FSI)
of the ejected nucleons were treated by optical potentials with parameters
generally obtained by fitting p-(A$-$1) elastic scattering data.  
Some of these calculations include also contributions to the
cross-sections from two-body currents such as meson-exchange currents
(MEC) and isobar configurations (IC).  DWIA calculations generally describe
well the shape of the measured distorted momentum distributions of valence 
states for missing momenta up to $250-300\,\mathrm{MeV/c}$,
but the calculated magnitudes are higher than those measured.
For any specific hole state, the ratio between the measured and the
calculated  cross-section gives the occupancy (or spectroscopic factor)
of the state in comparison  to mean-field single-particle models predictions.
For nuclei with $A>4$, for which mean-field calculations have been used,
the measured occupancy (see Fig.~\ref{fig:sfa}) is approximately
0.6~\cite{lapikas1}.

The low occupancy has been attributed to shifted strength from
single-particle states to more complicated configurations and
nucleon-nucleon correlations.  For missing momenta higher than 300 MeV/c,
as well as for high missing (excitation) energies, DWIA under-predicts
the measured cross-sections, supporting this hypothesis.

\medskip

In the last several years, new experimental facilities and theoretical advances
have come into being.  High-quality continuous-wave (CW) beams as well 
as high-resolution spectrometers provide very detailed and precise new data 
up to transfered momenta above 1 (GeV/c)$^2$.  Theoretically, fully 
relativistic DWIA calculations are now available for mean-field 
approaches to many-body nuclei~\cite{udias}, while microscopic
calculations with increasingly more relativistic ingredients 
are used for few-body systems~\cite{sabin1,sabin2,sabin3}.  It is 
possible now, therefore, to study nuclei and nuclear reactions
in great detail.  Thus, response functions and a
variety of kinematics provide selectivity in the contributions due
to various currents.  We are in a position, then, to constrain our
models and explicitly test their ingredients.

\medskip

This manuscript is written in light of these developments.
We do not attempt to provide a review of all \eep experiments.
Rather, we examine the promise and limitations in using them to study both 
nuclear structure and nuclear reactions.  We limit ourselves here 
to study of nuclei (i.e. $A\ge 2$).
Furthermore, although polarization observables are now routinely 
used in  \eep reactions, we deal here with reactions for which the 
only polarization observable is the beam-helicity asymmetry.

\section{\lowercase{\large (e,e$'$p)} on complex nuclei}

\begin{figure}[htb]
    \includegraphics[height=0.35\textheight]{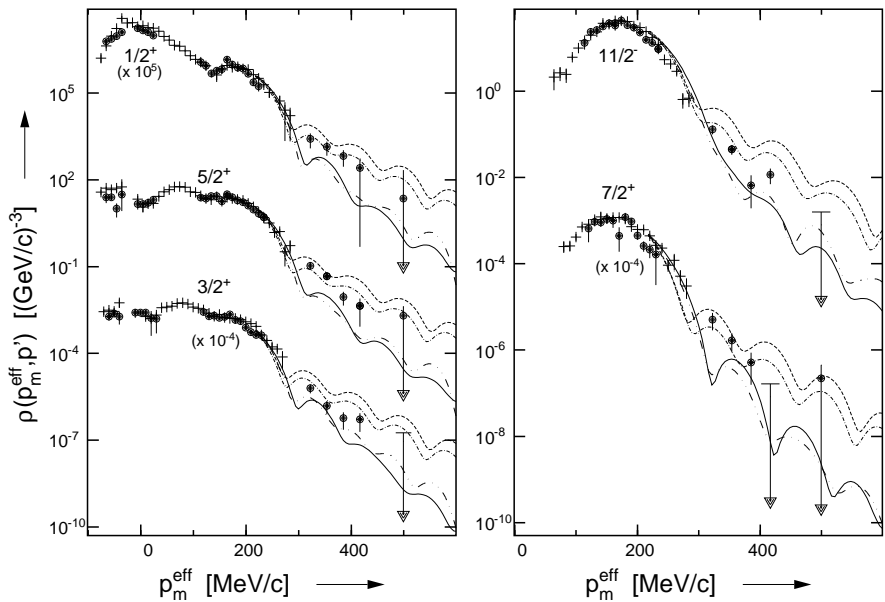}
    \caption{Missing momentum distributions for low-lying states in the
    reaction $^{208}$Pb(e,e$'$p)$^{207}$Tl.  Solid circular data are
    from Bobeldijk \etal~\protect\cite{bobeldijk1},
    and the plus data are from Quint \etal~\protect\cite{quint}.  
    The solid lines are DWIA calculations.
    Calculations that include correlations are by 
    Pandharipande~\protect\cite{vijay} (dash-double-dotted),
    Ma and Wambach~\protect\cite{ma} 
    (dashed) and Mahaux and Sartor~\protect\cite{mahaux1} (dot-dashed).}
\label{fig:Pb}
\end{figure}

A beautiful example of the success and limitations of previous studies
of complex nuclei is illustrated in Fig.~\ref{fig:Pb} from
reference~\cite{bobeldijk1}.  The missing-momentum distributions of
low-lying states in the reaction $^{208}$Pb(e,e$'$p)$^{207}$Tl that
have been measured in two experiments at NIKHEF are compared to several 
non-relativistic DWIA calculations.
The low p$_m$ data were measured in parallel kinematics,
whereas the data for 300 MeV/c $\le$ $p_\mathrm{m}$ $\le$ 500 MeV/c
were measured  at constant electron kinematics.
Since the purpose of the measurement was to look for correlations in the
ground-state wave function, most calculations~\cite{vijay,ma,mahaux1} 
used a variety of prescriptions to account for long-range and short-range
correlations (SRC). 

As can be seen, all DWIA calculations
reproduce the data well up to $p_\mathrm{m}$ = 300 MeV/c.
However, only qualitative
conclusions could be drawn as to the apparent effect of long-range 
correlations at low excitation energies, demonstrating the need for
additional observables to constrain the models. A case in point is 
a subsequent analysis of the same data~\cite{udiaspb} in which
fully relativistic mean-field DWIA calculations without correlations
have been used to consistently 
account for the observed missing-momentum distributions below and
above 300 MeV/c.
In what has been a general trend, the spectroscopic factors extracted from 
the relativistic analysis are higher
than those extracted from the non-relativistic calculations.  
Following the suggestion of M\"uther and Dickhoff~\cite{muther} that
short-range and tensor correlations should be present at 
high missing momenta and high excitation energies, further measurements 
were performed in this kinematic region~\cite{bobeldijk2}.

\medskip

\begin{figure}[htb]
    \includegraphics[height=0.35\textheight]{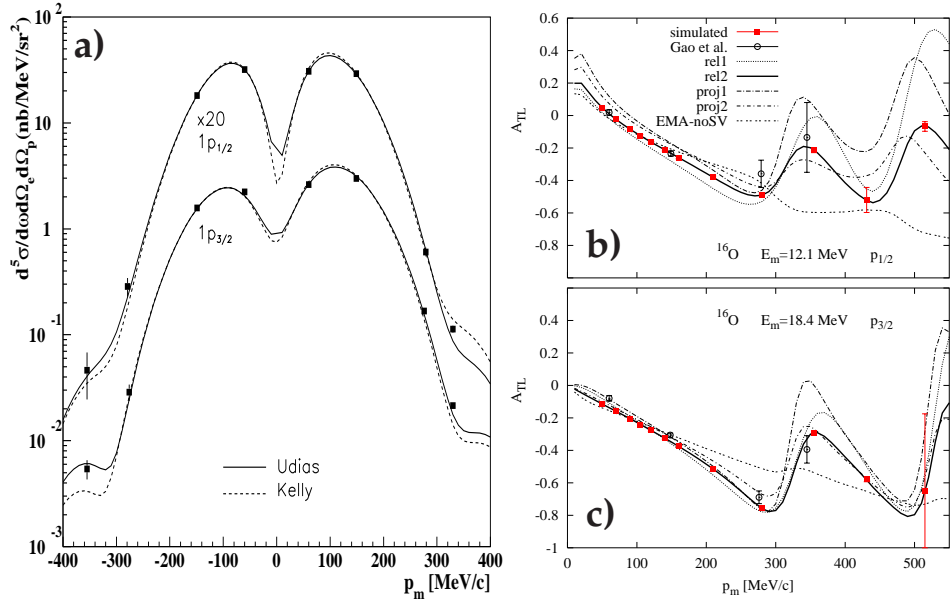}
    \caption{a) Measured cross-sections by Gao \etal~\cite{gao} and DWIA
       calculations by Udias \etal~\cite{udias1} and Kelly~\cite{kelly1}
       of the \OeepN reaction.  
    b) Measured and calculated A$_{LT}$ asymmetry for the 1$p_{1/2}$ state.
       See text for details.  c) Same as b) but for the 1$p_{3/2}$ state.}
\label{fig:gao}
\end{figure}

In a more recent experiment, J. Gao \etal~measured the reaction 
\OeepN~\cite{gao}.  Cross-sections to the two valence 1$p$-shell states, 
as well as response functions were measured in quasielastic kinematics
and Q$^2$ = 0.8 (GeV/c)$^2$ at the Jefferson Laboratory (JLab).
In Fig.~\ref{fig:gao}a the measured cross-sections to the two
1$p$-shell states up to $p_\mathrm{m}$ = 350 MeV/c are presented together
with fully relativistic DWIA calculations by Udias \etal~\cite{udias1}
and ``relativized'' DWIA calculations by Kelly~\cite{kelly1}.  The
data were measured at a constant electron kinematics and are very
precise up to the highest missing momentum.  
Both calculations use NLSH bound-state wave functions.  However,
while Udias solves the Dirac equation directly, the calculations
by Kelly use the effective momentum approximation
for the lower components of the Dirac spinors.  It is evident that
the fully relativistic calculations better describe the data at
the highest missing momenta.  The spectroscopic factors are 0.73 and 0.72
for the 1$p_{1/2}$ and 0.71 and 0.67 for the 1$p_{3/2}$ state for
the Udias and Kelly calculations, respectively.  The somewhat
higher spectroscopic factors than the ones in Fig.~\ref{fig:sfa}
are a common feature of relativistic calculations, and
suggest that a larger component of the wave function can be understood
in terms of relativistic mean-field models.

The importance of dynamical relativistic effects is further illustrated in 
Fig.~\ref{fig:gao}b and Fig.~\ref{fig:gao}c.  Fig.~\ref{fig:gao}b 
shows the measured and calculated A$_{LT}$ asymmetry for the
1$p_{1/2}$ and Fig.~\ref{fig:gao}c shows the same for the 1$p_{3/2}$ state.
All calculations are by Udias.  The rel1 and rel2 are the fully
relativistic calculations that use de Forest~\cite{deforest} cc1
and cc2 current operators respectively, the ``proj'' indicate that 
momentum-dependent positive energy projection operator was employed,
while the EMA-noSV indicates that the effective momentum approximations
was used and no spinor distortions were permitted.  The rel2 calculations
are the same as in Fig.~\ref{fig:gao}a. It is clear from the figures
that spinor distortions arising from the enhancement of the lower
components due to the scalar and vector potentials are necessary to
adequately describe the structure of the observed A$_{LT}$ asymmetries.
The R$_{LT}$, R$_L$ and R$_T$ which were measured by
Gao \etal~are reproduced as well \cite{gao} by Udias' calculations,
whose only adjustable parameter was the spectroscopic factor.  It strongly
suggests that in these calculations, we have a successful model for
single-particle description of \AeepB reactions on complex nuclei.
That said, there is a need to test the limits of this model for high
missing momenta, and the ``simulated data points'' (full squares) in
Figs.~\ref{fig:gao}b,c are predicted values of the A$_{LT}$ asymmetries
up to missing momenta of 550 MeV/c using the same model by Udias.
These, as well as the cross-sections to $p_\mathrm{m}$ = 700 MeV/c will be 
measured at JLab in 2001~\cite{e00102}.
It is noteworthy that work is in progress to include MEC into these 
\AeepB relativistic calculations~\cite{amaro1,amaro2}.
Other work is being done  to consolidate the use of optical potentials
with that of the eikonal approximation for the treatment of
FSI~\cite{ryckebusch1}.

\begin{figure}[htb]
  \begin{center}
    \includegraphics[height=0.40\textheight]{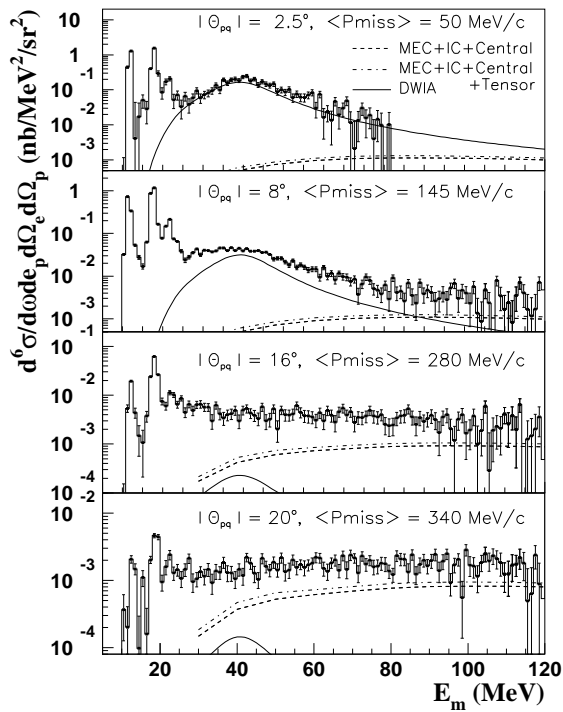}
    \caption{\label{fig:nilanga}
    Cross-sections as functions of missing energies for different
    missing momenta.  DWIA are calculations by Kelly~\cite{kelly1}
    in which the 1$s$-shell strength is folded with the Lorentzian
    parameterization by Mahaux.  The dashed and dot-dashed curves are
    calculations by Ryckebusch \etal~\cite{ryckebusch2} of the
    contributions to the \Oeep
    cross-section from \Oeenp and \Oeepp. (See text for details.)}
    \end{center}
\end{figure}

It is already evident that \Aeep reactions on complex nuclei to high
missing (excitation) energies provide only qualitative information,
and mainly on the reaction mechanism.  A recent measurement of the \Oeep
reaction~\cite{nilanga} at JLab at Q$^2$ = 0.8 (GeV/c)$^2$ clearly
illustrates this point.  The cross-sections for missing energies
ranging from the 1$s_{1/2}$ hole state to 120 MeV
were measured at missing momenta of 50-340 MeV/c.  The data are presented
in Fig.~\ref{fig:nilanga} and compared to the same single-particle
``relativized'' DWIA calculations by Kelly~\cite{kelly1}  that are reasonably
successful for the 1$p$ states (see Fig.~\ref{fig:gao}a).
The spectroscopic factor was 0.73.  It is strikingly clear that a
proton knockout from the 1$s_{1/2}$ shell accounts for the cross-section
at $p_\mathrm{m}$ = 50 MeV/c, but  the contribution to the cross-sections
from 1$s_{1/2}$ knockout decreases fast with increasing
missing momentum, and at $p_\mathrm{m}$ = 200 MeV/c, it is less than
10 percent. Above this $p_\mathrm{m}$, the cross-sections are
approximately flat for E$_m > 25$ MeV.  

\medskip

Also presented in the figure are calculations in a Hartree-Fock
framework by Ryckebusch \etal~\cite{ryckebusch2} of the 
contributions to the \Oeep cross-section
from \Oeenp and \Oeepp due to MEC, IC and central (Jastrow)
and tensor correlations.  All these contributions can 
only account for about a half of the observed cross-sections 
at high missing momenta.  In these calculations,
two-body (MEC and IC) currents constitute about 85\% of the calculated
\Oeepp and \Oeenp contributions, tensor short-range correlations
are about 13\% and central SRC are only 2\%.
The large contribution from two-body currents is consistent with the
measured response functions that were separated by 
Liyanage \etal~\cite{nilanga} as well.  The cross-sections at high
missing energies are more transverse than predicted by DWIA for
single-particle states, and increasingly so with increasing 
missing momentum.  Excess transverse cross-sections at high $E_\mathrm{m}$
had been previously measured for $^{12}$C by Ulmer \etal~\cite{ulmer}
at Q$^2$ = 0.15 (GeV/c)$^2$ and Dutta \etal~\cite{dutta} at 
Q$^2$ = 0.64 and 1.8 (GeV/c)$^2$.
Dutta's observation that the excess transversity decreases with Q$^2$
is consistent with Liyanage's results as well.

\medskip

Although a qualitative picture emerges from studying the
dynamics of \Aeep reaction at high missing energies, models are
far from being consistent and quantitative because the task of
disentangling reactions effects (FSI, MEC, IC etc.) from contributions
to the wave function from N-N correlations is too complex for the \eep
reaction alone. Hopefully, data from other
reaction channels such as A\eepp, A\eenp and A\eex, where X is any
combination of hadrons, can provide qualitative and quantitative knowledge
on the channels contributing to the \Aeep reaction.  
With such input, it may be possible to shed light on the elusive
subject of short-range correlations.
Such experiments are being now performed, but they are outside of
the scope of this manuscript.

\section{\lowercase{\large (e,e$'$p)} on few-body systems}

From a strictly experimental point of view, the use of \Aeep reactions
to study few-body nuclear systems is not different from the study of
complex nuclei.  However, there are differences in the
theoretical treatments that affect the present and future
possibilities and limitations associated with these reactions.

\begin{figure}[htb]
    \includegraphics[height=0.30\textheight]{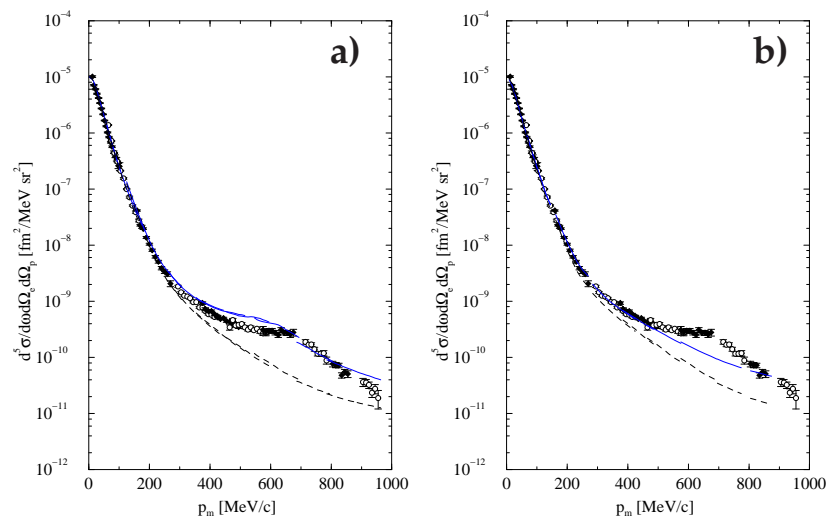}
    \caption{Measured and calculated d(e,e$'$p) cross-sections.
    In both panels, the solid curves
    are calculations that include MEC and IC, and the dashed curves
    are calculations that do not include contributions from two-body
    currents.  a) Calculations by Arenh\"ovel (private communications
    in~\cite{blomqvist1}).  b) Calculations using a code by Schiavilla
    (private communications in~\cite{blomqvist1}.)}
\label{fig:Mainzd}
\end{figure}

The electrodisintegration of the deuteron, \deep, has been used to
study N-N potentials, the structure of the deuteron and reaction dynamics.
Fig.~\ref{fig:Mainzd} presents results from a recent experiment
at MAMI~\cite{blomqvist1}.  Cross-section over a range of 6 orders of
magnitude were measured with high precision as a function of missing 
(recoil) momentum up to $p_\mathrm{m}$ = 950 MeV/c.  Because of the low
beam energy of the MAMI microtron, this measurement could not be performed
at a constant electron kinematics, hence imposing ambiguities on the
theoretical interpretation of the data.  Nevertheless, such a measurement
that was impossible only ten years ago, provides invaluable and badly 
needed input to reaction models.  The higher beam energy available at JLab
makes it possible to perform a similar measurement at
a constant electron kinematics, eliminating some ambiguities from
the theoretical interpretation.

\medskip

Also presented in the two panels of Fig.~\ref{fig:Mainzd} are two
state-of-the-art calculations by Arenh\"ovel and Schiavilla,
each with and without MEC and IC.  In both calculations the clear
importance of including these two-body currents for reproducing even
the gross features of the cross-sections  above p$_m = 300$ MeV/c is
evident.  Since these high $p_\mathrm{m}$ data were measured
at $\Delta$ kinematics, the contributions from IC are more important
than those from MEC.  Yet, the two calculations are very different
in treating the two-body currents and produce different calculated cross
sections.  This is a clear indication that a measurement of a single
observable (cross-sections in this case) is insufficient to fully
disentangle the contributing ingredients.  As many observables as
possible, and at more than one kinematics are necessary for this purpose.

\begin{figure}[htb]
  \begin{center}
    \includegraphics[height=0.37\textheight]{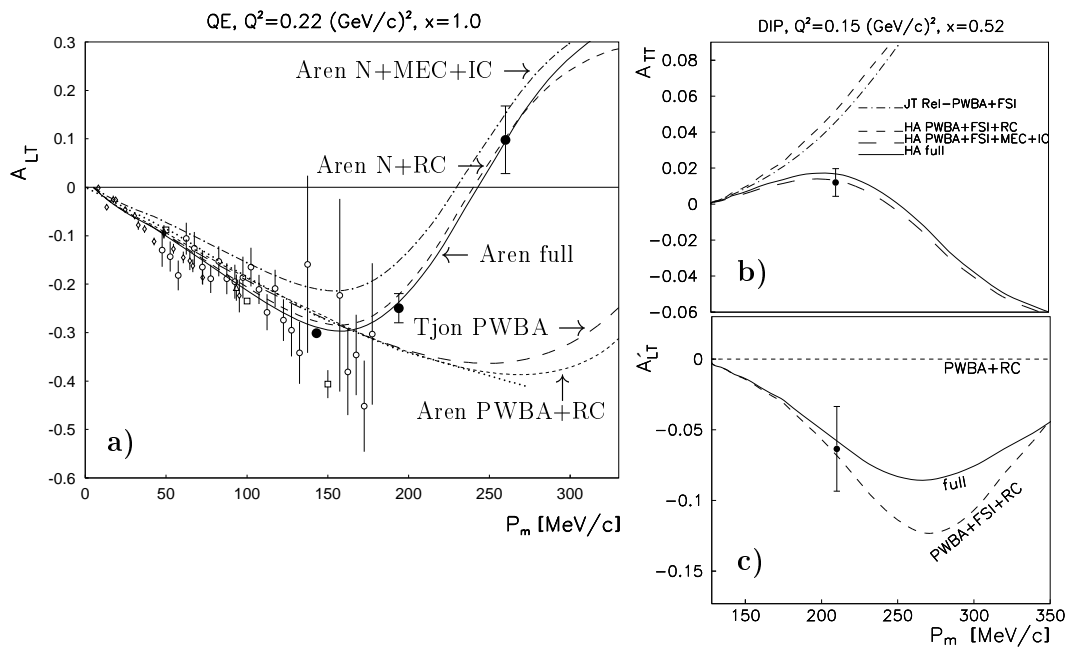}
    \caption{\label{fig:oopsd}
    Interference asymmetries for the \deep reaction.  The solid circles
    are data from reference~\cite{zilu}.  The open circles are previous
    data. The calculations by Tjon are fully relativistic.
    The calculations by Arenh\"ovel are based on the Schr\"odinger equation
    and include relativistic corrections.  The dotted curve is the
    cc1 current operator as prescribed by de Forest~\cite{deforest}.
    a) A$_{LT}$ in quasielastic kinematics.
    b) A$_{TT}$ in dip kinematics.  c) A$'_{LT}$ in dip kinematics.}
    \end{center}
\end{figure}

An example of an attempt to disentangle the model ingredients in the
\deep reaction is the program at the MIT-Bates Linear Accelerator Center
to measure the four helicity-independent unpolarized response functions
R$_L$, R$_T$, R$_{LT}$, R$_{TT}$ and the (``fifth'') 
helicity-dependent response R$'_{LT}$.  Since the two latter responses
require the detections of the knocked-out proton out of the
electron-scattering plane, a system was constructed of four
out-of-plane magnetic spectrometers (OOPS)~\cite{oops} that 
can be positioned around the momentum-transfer vector, each with a
precision better than $\pm 0.3$~mm and $\pm 0.3$~mrad.  Initial results
at Q$^2$ = 0.22 (GeV/c)$^2$~\cite{zilu} of the interference A$_{LT}$
asymmetry at quasielastic kinematics and at Q$^2$ = 0.15 of the 
interference A$_{TT}$ and A$'_{LT}$ asymmetries in dip kinematics
are presented in Fig.~\ref{fig:oopsd} together with calculations
by H. Arenh\"ovel and J. Tjon.  The sensitivity of each response
to the various ingredients is clearly demonstrated in the figure.
A$_{LT}$ is very sensitive to relativistic effects (as already noted
above for \OeepN).  Yet, since Tjon's relativistic calculations
do not include contributions from FSI,  they fail to reproduce
the data.  A$_{TT}$ is sensitive mainly to two-body currents,
whereas A$'_{LT}$ is mainly sensitive to FSI.  Furthermore, dip
kinematics enhances the contributions from two-body currents,
while $\Delta$ kinematics is expected to emphasize IC over
MEC.  Hence, precise and simultaneous measurements of all these
responses at the various kinematics constrain the models to the
point that all ingredients have to be treated correctly.  
In the initial results of Fig.~\ref{fig:oopsd}, the calculations
by Arenh\"ovel are able to reproduce all the data.  It remains to be seen
whether these calculations are able to reproduce future data at a higher
$p_\mathrm{m}$ and at $\Delta$ kinematics, from measurements that are
planned at Bates~\cite{batesd}.

\medskip

The R$_L$ and R$_T$ responses have been measured using the ``Rosenbluth
separation'' method, whereby the cross-sections are measured in parallel
kinematics for constant transfered  momentum and energy at two electron
angles (virtual photon polarizations).  At $p_\mathrm{m}$ = 50 MeV/c,
results from Saclay and Bates are not consistent with those from NIKHEF
~\cite{jordan} in that their measured R$_L$ is lower by about 40\%.
Moreover, their measured R$_L$ response is about 20\% lower than
predicted by Arenh\"ovel's calculations.  Preliminary data
from a more recent measurement at Mainz~\cite{werner1} also indicates
a smaller R$_L$ than predicted.
This poses a huge problem, since there is very little room in the
theory to adjust the R$_L$ response.
It has been suggested that the discrepancy between the measured
and predicted R$_L$ may be due to
the fact that measurements in parallel kinematics are prone to 
large uncertainties due to the correlation between $p_\mathrm{m}$ and the
transfered energy, and at low $p_\mathrm{m}$
also due to ambiguities in the definition of the $\vec{p}_m$ vector.
Consequently, an alternative 
``Rosenbluth separation'' was proposed at Bates using perpendicular
kinematics. Although in 
perpendicular-kinematics ``Rosenbluth separation''  R$_L$ is contaminated
by a small contribution from R$_{TT}$,  this R$_{TT}$ contribution can
be directly and  simultaneously measured using OOPS, providing a clean
and alternative measure of the R$_L$ response.

\begin{figure}[htb]
    \includegraphics[height=0.5\textheight]{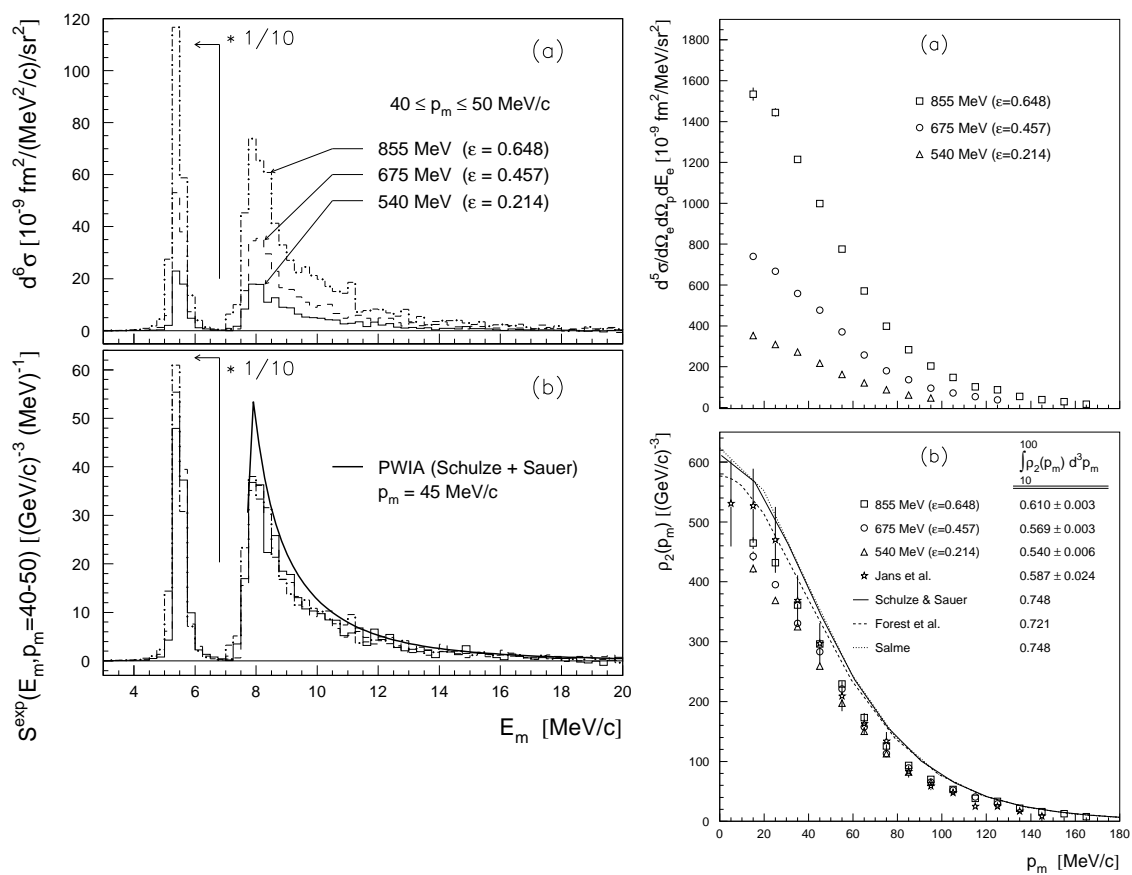}
    \caption{Left panel: a) {\mbox{$^{3}$He(e,e$'$p)}} cross-sections
    and b) experimental (``distorted'') spectral functions of missing 
    energy for a missing momentum
    bin of  40-50 MeV/c.  Right panel: a) {\mbox{$^{3}$He(e,e$'$p)$^2$H}}
    cross-sections as a function of missing momentum. b) extracted 
    (``distorted'') momentum
    distributions for the two-body breakup process compared to various
    calculations.   Figures b (left and right)
    were obtained by dividing out the electron-proton cross
    sections from the data in figures a (left and right), respectively.
    The figures are from reference~\cite{richard}.}
\label{fig:Mainz3he}
\end{figure}

Data are also available on {\mbox{$^{3}$He(e,e$'$p)}}
and {\mbox{$^{4}$He(e,e$'$p)}}.  The cross-sections and the R$_L$, 
R$_T$ and R$_{LT}$ responses were measured for both nuclei.  However,
by and large, they were measured in sporadic kinematics, making a
consistent theoretical interpretation difficult.  Nevertheless, it
is worth mentioning the example of a program at MAMI~\cite{mamihe34}
to measure the R$_L$ and R$_T$ responses for
{\mbox{$^{3}$He(e,e$'$p)}} and {\mbox{$^{4}$He(e,e$'$p)}}
in the same transfered momentum and for three transfered energies on,
above and below the quasielastic peak.  The idea is to selectively
emphasize the single particle aspects of the reaction in quasielastic
kinematics, MEC and IC in dip kinematics and SRC below the quasielastic
peak (x$_B > 1$).  The data for {\mbox{$^{3}$He(e,e$'$p)}} in
quasielastic kinematics have been published~\cite{richard} with the
main conclusions that the L/T behavior of
the cross-section follows that of the electron-proton's, and that MEC
and IC do not play a noticeable role in this cross-section.
Fig.~\ref{fig:Mainz3he} demonstrates the first conclusion.

\medskip

Results for the quasielastic {\mbox{$^{4}$He(e,e$'$p)}} measurement
are soon to be submitted for publication and show similar L/T behavior,
as do results for {\mbox{$^{3}$He(e,e$'$p)}} at x$_B > 1$~\cite{sasha}.
For both nuclei, the R$_{LT}$ response was also measured at MAMI for
a variety of transfered momenta and the results will be published
upon completion of the analysis.

\medskip

With the availability of high-energy, high quality beams  and
detectors with large acceptance and high precision it is possible now
to look for the transition from hadronic to quark-gluon (or quark and
flux tubes) degrees of freedom in the nucleus.  The idea is to reach
the limit of being able to describe a large data set with 
``conventional'' nuclear physics.  For this purpose it is necessary,
and work is underway,  to
develop what theorist refer to as a ``standard model of nuclear
physics'', that includes a consistent treatment of FSI, two-body currents,
and relativity (wave-functions, current operators and dynamics).  
Because of the deuteron's simplicity, the \deep reaction at high energies
and transfered momenta can be used as a unique laboratory for
this study.  A program is pending at 
JLab~\cite{werul} to test the models that are being developed by 
measuring the cross-sections and responses of the \deep 
reaction at high energies and transfered momenta and at a
variety of kinematics.  However, such studies cannot be confined
to the deuteron, if a consistent theory is to be developed.
A similar program for {\mbox{$^{3}$He(e,e$'$p)}} is already
underway~\cite{89044}, with preliminary data to be presented in
the Fall 2001 DNP meeting in Hawaii.  A proposal for
{\mbox{$^{4}$He(e,e$'$p)}} has been approved by the JLab PAC as
well~\cite{01108}.  Data on $^4$He will have an added importance,
as they can provide a bridge between microscopic and mean-field
\Aeep calculations.  With the above in mind, the immediate limitations
of \eep experiments at high energies and transfered momenta will be due
to theoretical interpretation.  In turn, theoretical progress can only
be made if high quality data on a rich set of observables
are available to constrain and test it.

\section{Summary}

\eep is a powerful tool for studying nuclear structure and interactions.
High quality and consistent data on many observables and at a variety of
kinematics are needed to constrain and test theoretical
models.  Fortunately, we are at a stage where such data are feasible
to obtain, and indeed are beginning to emerge.  Simultaneously,
theoretical models are being developed in which relativity, modern
approaches in treating correlations, FSI and two-body currents
as well as sophisticated computational techniques play
increasingly important roles.

\medskip

For many-body nuclei relativistic mean-field DWIA calculations are very 
successful in predicting cross-sections and responses in \AeepB up to 
missing momenta of 350 MeV/c.  Measurements are pending to 
determine the $p_\mathrm{m}$ limits of this model, while
work is also being done to include two-body currents in this model.  For
higher residual excitations, the usefulness of \Aeep reactions is limited
by the complexity of channels contributing to the reaction.  Such reaction
channels must be measured explicitly in order to provide additional input
for the analysis of the \Aeep channel.  It is anticipated that the
next-generation MAMI accelerator, with its high-quality 1.5 GeV beam will
play a central role in the study of complex nuclei.

\medskip

Theories are developing faster in the few-body sector.
Exact non-relativistic calculations are available for the deuteron and
$^3$He, and variational Monte Carlo techniques are used for $^{3,4}$He.
Two-body currents are treated in all calculations, including the
Faddeev-based calculations for the three-nucleon system.  Relativity
is included to various degrees in all calculations, either as corrections
or explicitly.  Relativistic current operators and, in some cases,
wave functions are increasingly included in these calculations.

\medskip

With the availability of high energy and high quality beams at CEBAF,
the extension of these studies to high transfered momenta and energies
can be pursued.  Here there is the promise of reaching the transition
from hadronic to quark-gluon (or quark and flux-tubes) degrees of freedom.
In order to recognize this transition, a consistent ``standard nuclear
model'' for the hadronic degrees of freedom is needed, 
the breakdown of which will indicate the emergence of the new degrees
of freedom.  This can be achieved only by providing a large experimental
data set with many observables, at various kinematics and on several
nuclei to constrain and test the developing models.

\section*{Acknowledgments}

\noindent I would like to thank S. \v Sirca and Z.-L. Zhou for their 
assistance in preparing this manuscript.


\end{document}